\documentclass{aa}

\usepackage{txfonts}
\usepackage{graphicx}
\usepackage{amssymb}
\usepackage[below]{placeins}
\usepackage{natbib}
\bibpunct{(}{)}{;}{a}{}{,}

\begin{document}

\title{X-ray spectroscopy on Abell 478 with XMM-Newton}

\author{J. de Plaa \inst{1,2} \and 
	J. S. Kaastra \inst{1} \and 
	T. Tamura \inst{1,3} \and 
	E. Pointecouteau \inst{4} \and
	M. Mendez \inst{1} \and 
	J. R. Peterson \inst{5} }

\institute{SRON National Institute for Space Research, Sorbonnelaan 2, 3584 CA Utrecht, The Netherlands\\
           \email{j.s.kaastra@sron.nl}
	   \and Astronomical Institute, Utrecht University, PO Box 80000, 3508 TA Utrecht, The Netherlands
           \and Institute of Space and Astronautical Science, JAXA, 3-1-1 Yoshinodai, Sagamihara, Kanagawa 229-8510, Japan
	   \and CEA/DSM/DAPNIA Saclay, Service d'Astrophysique, L'Orme des Merisiers, B\^at. 709, 991191 Gif-sur-Yvette, France
	   \and KIPAC, Stanford University, PO Box 90450, MS 29, Stanford, CA 94039,USA
	   }

\offprints{J. de Plaa, \email{j.de.plaa@sron.nl}}

\date{17 May 2004}

\abstract{
We report the results from a spatially resolved spectroscopy study with XMM-Newton on the relaxed cluster of galaxies \object{Abell 478}.
From the EPIC data we extract a temperature profile and radial abundance profiles for Ne, Mg, Si, S, Ca, Fe and Ni. The abundance profiles
follow the same trends as observed in other clusters.  
The spectra of the core of the cluster can be best fitted with a multi-temperature model. We argue that this 
multi-temperature behavior is mostly due to projection effects, because of the strong temperature gradient in the core. 
Contributions from other effects, for example, intrinsic temperature stratification cannot be fully excluded.
For the first time we measure an underabundance of oxygen in the Galactic absorption component 
toward a cluster. The measured oxygen abundance in this absorber is about 0.5 times the solar oxygen abundance as determined by \citet{anders1989}.

}

\maketitle

\section{Introduction}

The evolution and hydrodynamical structure of the hot diffuse X-ray emitting gas in clusters of galaxies is still not well
understood. Several clusters show excess X-ray emission in their cores. The so-called cooling flow models
\citep[for a review see][]{fabian1994} propose radiative cooling of gas in the core to explain the observed surface brightness 
and temperature profiles. The pressure decrease in the core due to the cooling gas causes a net inflow toward 
the center of the cluster, hence the name cooling flow. XMM-Newton observations show that the amount of cool gas 
in the core of cooling-flow clusters is much lower than predicted 
\citep{peterson2001,kaastra2001,tamura2001a,peterson2003,kaastra2004}. Therefore, the standard cooling-flow 
model needs to be adjusted. Several mechanisms to explain the observations have been proposed 
\citep[see][ for a list of possible explanations]{peterson2001}.

The \object{Abell 478} cluster of galaxies is a good example of a highly relaxed cluster. 
Earlier X-ray observations with EXOSAT \citep{edge1991}, Ginga and Einstein \citep{johnstone1992},
ROSAT \citep{allen1993,white1994}, ASCA \citep{markevitch1998,white2000}
and Chandra \citep{sun2003} show a significant excess of X-ray emission in the core, suggesting the presence of a 
massive cooling flow. The observations performed by missions with sufficient spatial resolution (e.g. ROSAT, 
ASCA and Chandra) are consistent with a radial temperature decrease toward the core. In a recent Chandra
observation an X-ray cavity was discovered in the core which seems to be associated with a lobed radio source \citep{sun2003}. 

In this paper we study the properties of \object{Abell 478} using high-resolution and spatially-resolved spectra obtained
with the European Photon Imaging Camera \citep[EPIC,][]{turner2001} and the Reflection Grating Spectrometer \citep[RGS,][]{herder2001} 
aboard XMM-Newton \citep{jansen2001}. We focus on the physics and metal abundances in the core of \object{Abell 478}. In an 
associated paper \citet{pointecouteau2004} describe the large scale temperature, gas and dark matter distribution in this cluster using a pointed 
and an offset XMM-Newton EPIC observation.  

Throughout this paper we use $H_{0}$ = 50 km s$^{-1}$ Mpc$^{-1}$ and q$_0$ = 0.5. Using this cosmology 1$\arcmin$ is 157 kpc at 
the cluster redshift of 0.0881.

\section{Observations and data analysis}

The observation of \object{Abell 478} was performed as part of the Guaranteed Time program on February 15 2002 and had a total duration of 126 ks.
Both EPIC MOS cameras were operated in Full Frame mode and the EPIC pn camera in Extended Full Frame mode. For all EPIC cameras 
the thin filter was used. 
 
For the analysis we use the event lists produced by the XMM-Newton SOC and the 5.4.1 version of the XMM Science Analysis System (SAS).  
To correct for enhanced and variable background due to soft proton flares we use the events with energies larger than 10 keV from 
pn and CCD 9 of RGS. The data are filtered using upper and lower count-rate thresholds. The 
thresholds are calculated in a way analogous to the method used in \citet{pratt2002}. We first make a histogram of the  
light curve above 10 keV with a binsize of 100 s and pattern equal to 0, which we fit it with a Poissonian function. The count rate 
thresholds are subsequently fixed to $N \pm 3\sqrt{N}$, where $N$ is equal to the mean number of counts within a bin. 
Because the pn is most sensitive to solar flares, we merge the Good Time Intervals (GTI) derived from the pn data also with the 
preliminary GTI for MOS. Therefore pn and MOS have roughly the same exposure time. The maximum count rates and the resulting useful 
exposure times are shown in Table~\ref{tab:exposure}. This simple variable background procedure aims at obtaining maximum statistics
in the core region, where the background is relatively less important, but excludes contributions from the large solar flares. 
Because the cluster signal in the 0--4$\arcmin$ region dominates over the background, it is not necessary to perform a double
background subtraction nor a multi-band flare search like described in \citet{pointecouteau2004}.
The countrate thresholds we chose are well within the advised threshold of 
$<$1.0 cts s$^{-1}$ (PN) provided by the XMM-Newton SOC \citep{kirsch2003}.

\begin{table}
\caption{Maximum count rates and resulting useful exposure times after screening for solar flares.}
\begin{center}
\begin{tabular}{lcc}
\hline\hline
Detector	  & Max. count rate  	     & Useful exposure time\\
		  & (counts s$^{-1}$) 	     & (ks)	\\
\hline
PN		  & 0.53		     & 51	\\
MOS		  &			     & 51	\\
RGS		  & 0.36		     & 119	\\
\hline		  
\end{tabular}
\end{center}
\label{tab:exposure}
\end{table}

\subsection{EPIC}
\label{sect:obs_epic}

Because the cluster is slightly elongated along the north-east to south-west axis, we extract 
the source and background spectra from elliptical annuli allowing patterns $\leq$12 for MOS and $\leq$4 for pn.
We use a ratio of 1.2 between the major and minor axis of the ellipse, which we obtain from an empirical 2D 
$\beta$-model fit to the EPIC image. Although the data show some minor deviations from this 2D model, the value is in 
the range of 1.2--1.4 \citep[ROSAT,][]{white1994} and compatible with 1.22 derived by \citet{pointecouteau2004} who uses the 
same method and data. 
Further on in this paper we characterize the extraction region by its semi major axis.   
We extract the EPIC background spectra from the blank sky event files provided by \citet{read2003}.
Relatively bright point sources in the EPIC field of view have been excluded both from the \object{Abell 478}
and background datasets.  

To gain statistics in each spatial bin we extract the spectra from annuli with a width of more than 30$\arcsec$.   
This way, we are less sensitive to the energy dependent shape of the PSF and therefore we neglect this effect
in the rest of the analysis. The downside of choosing large binsizes is that mixing of temperatures within the 
bin also increases with binsize if there are strong temperature gradients present. This effect may become 
important during multi-temperature fitting. 

Although we neglect the influence of the energy dependent part of the PSF, a substantial broadening effect 
in the core remains. To correct the normalization of each annulus for PSF effects we calculate correction 
factors by fitting the surface brightness profile from the Advanced Camera Imaging System (ACIS) on board 
Chandra \citep{sun2003} with a double King profile. We normalize both the ACIS fit 
and the EPIC surface brightness profile and calculate the ratios between the two for each bin. These
factors range from 0.73 in the core up to 1.03 at 4$\arcmin$.  
Then, we multiply the effective area of each annulus with this factor and obtain a PSF corrected response.

\begin{table}
\caption{Systematic errors included in the EPIC spectral fits. \citep[Adapted from][]{kaastra2004} }
\begin{center}
\begin{tabular}{lcc}
\hline\hline
Energy band  	& Error source 	& Error background \\
\hline
0.4--0.5	& 10\% 		& 35\% \\
0.5--0.7	& 5\%		& 25\% \\
0.7--2.0	& 5\%		& 15\% \\
2.0--10.0	& 5\% 		& 10\% \\
\hline
\end{tabular}
\end{center}
\label{tab:systematic}
\end{table}

For the spectral analysis we use the SPEX package \citep{kaastra2003} and fit the spectra over the 0.4--10 keV range. 
We include systematic errors due to uncertainties in the calibration and the background in the spectral fit. The applied 
systematic errors are shown in Table~\ref{tab:systematic}; they have been adapted from \citet{kaastra2004}. 
These systematic errors enclose possible variations of the Cosmic X-ray Background (CXB), background normalization
and errors in the calibration.  

After a preliminary analysis, the pn appears to have a gain problem, resulting in a significantly lower value for the 
redshift of the source than measured by the MOS instrument. 
From MOS we obtain a redshift of 0.0889 $\pm$ 0.0003, which is consistent with the optical value of
0.0881 $\pm$ 0.0009 \citep{zabludoff1990}. But, our fits of the pn spectra result in a redshift of 0.0775 $\pm$ 0.0002.  
To correct for this, we measure the centroid energies of the aluminum, nickel,
copper and zinc background lines. A linear fit ($E_{\mathrm{new}} = aE_{\mathrm{old}} + b$) through the measured and expected energies for these lines yields 
$a =$ 1.0066 $\pm$ 0.0007 and $b =$ (-1.04 $\pm$ 0.24) $\times$ 10$^{-2}$. After correcting
the energy values in the original eventfile according to this relation, we fit the resulting spectra again. 
The corrected spectra show large discrepancies at low energies. Therefore, we conclude that the observed gain problem is 
probably not linear, in line with the argumentation of \citet{pointecouteau2004}. 
Unfortunately, the number and distribution of background lines is insufficient to derive the exact shape of the gain
correction. For the rest of the analysis we therefore use the original event file, fit MOS and pn 
separately and let the redshift free in the fit.

\begin{figure*}
\begin{minipage}{0.33\textwidth}
\includegraphics[width=1.0\textwidth]{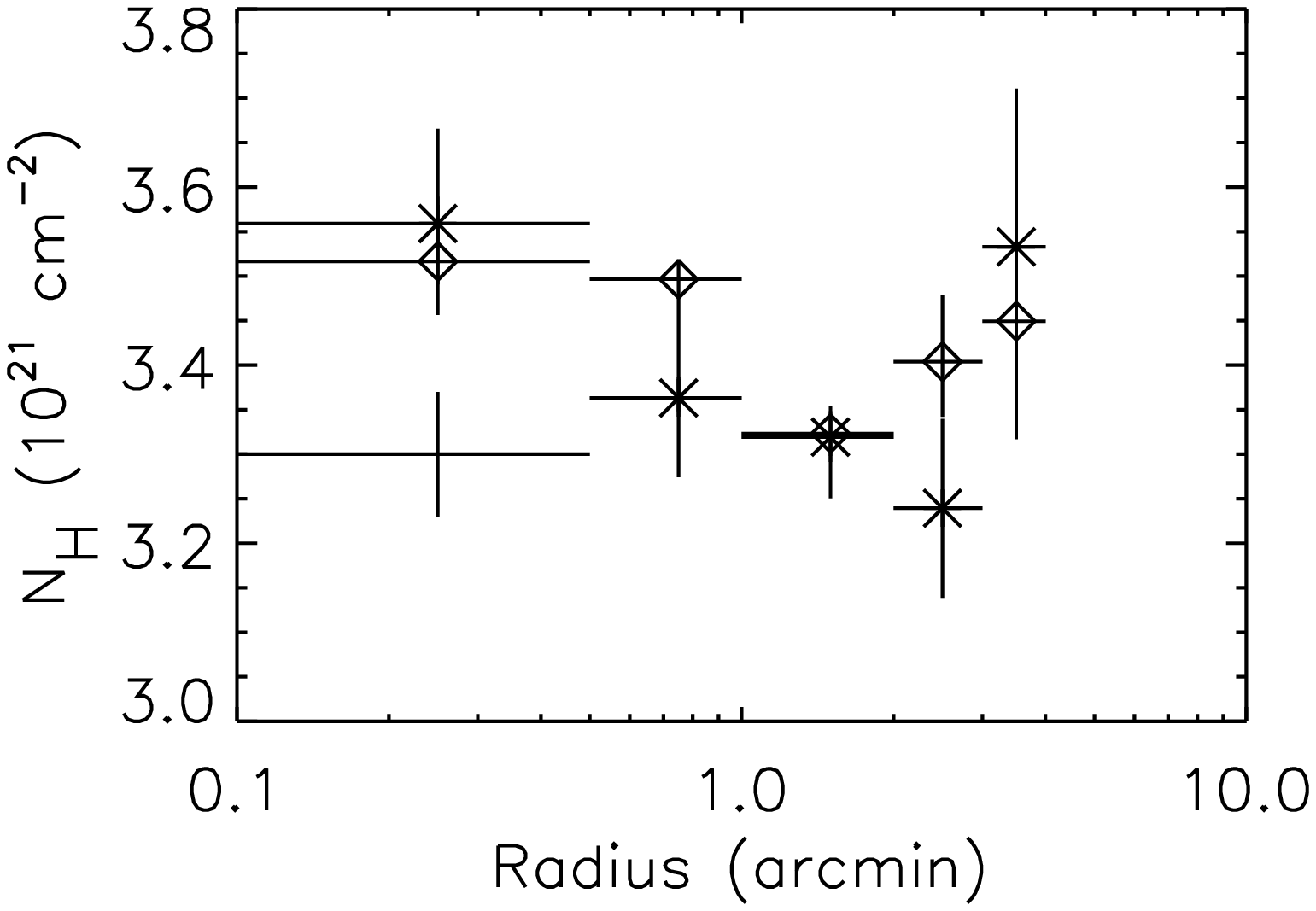}
\end{minipage}
\begin{minipage}{0.33\textwidth}
\includegraphics[width=1.0\textwidth]{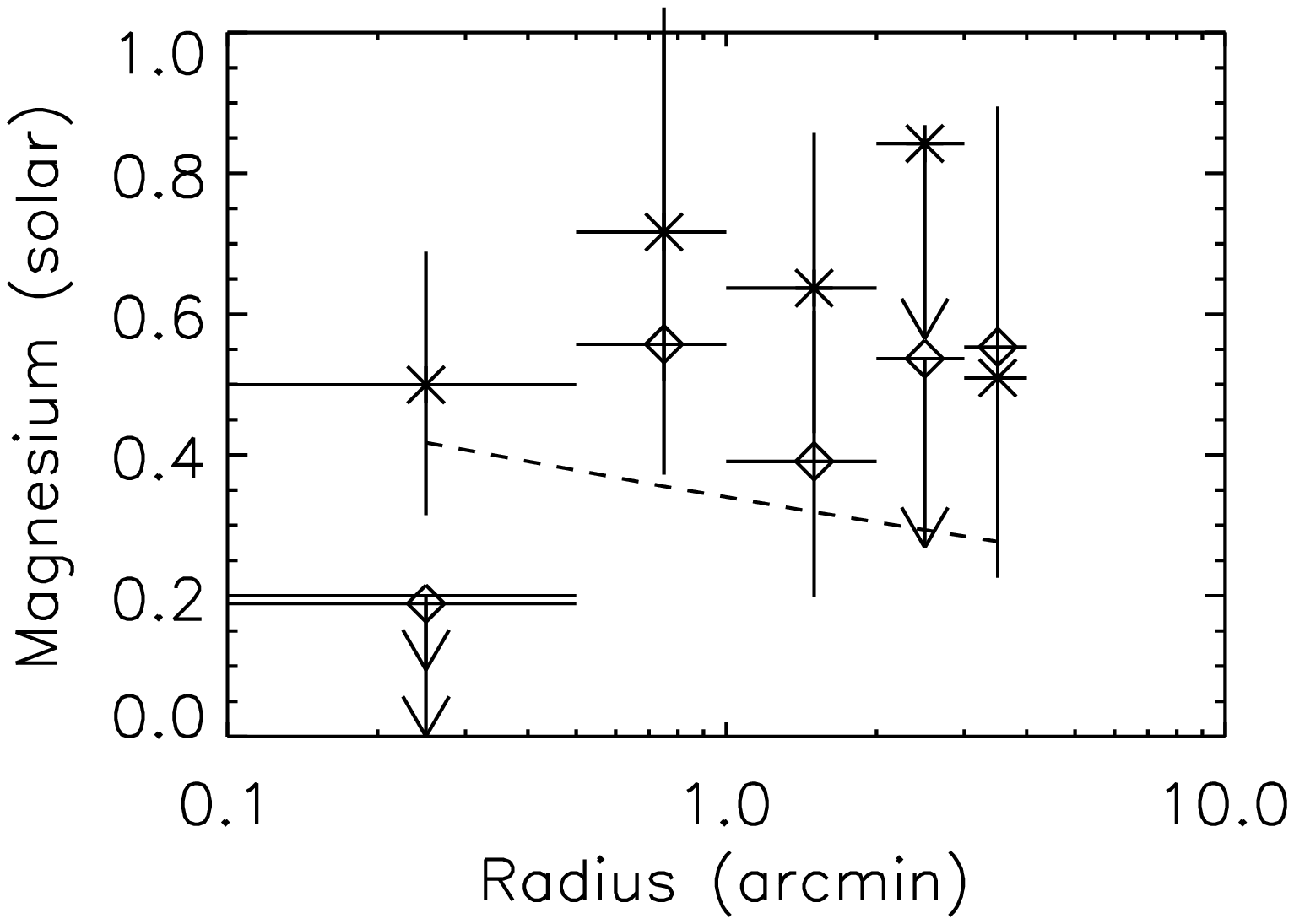}
\end{minipage}
\begin{minipage}{0.33\textwidth}
\includegraphics[width=1.0\textwidth]{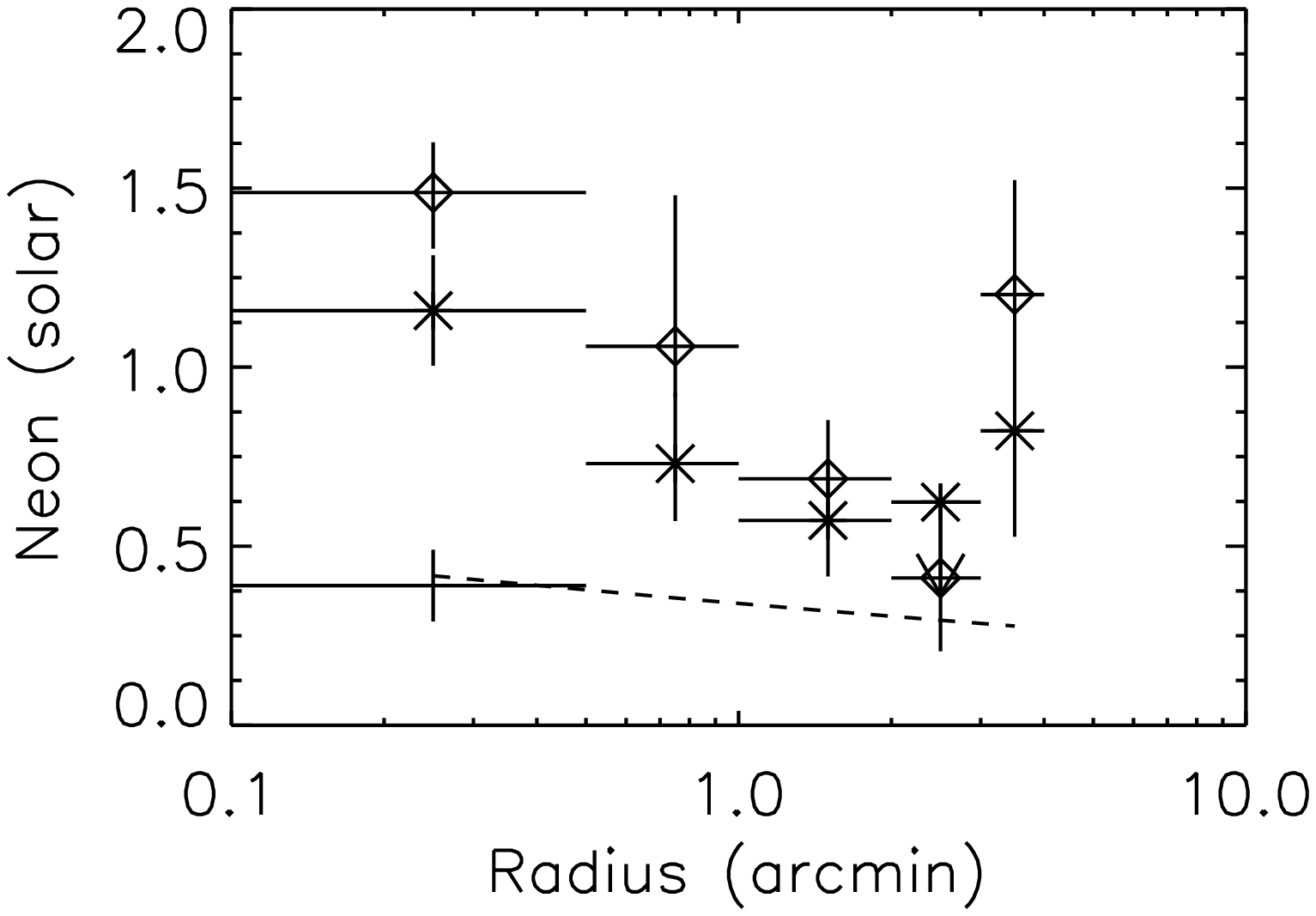}
\end{minipage}\\
\vspace{2mm}

\begin{minipage}{0.33\textwidth}
\includegraphics[width=1.0\textwidth]{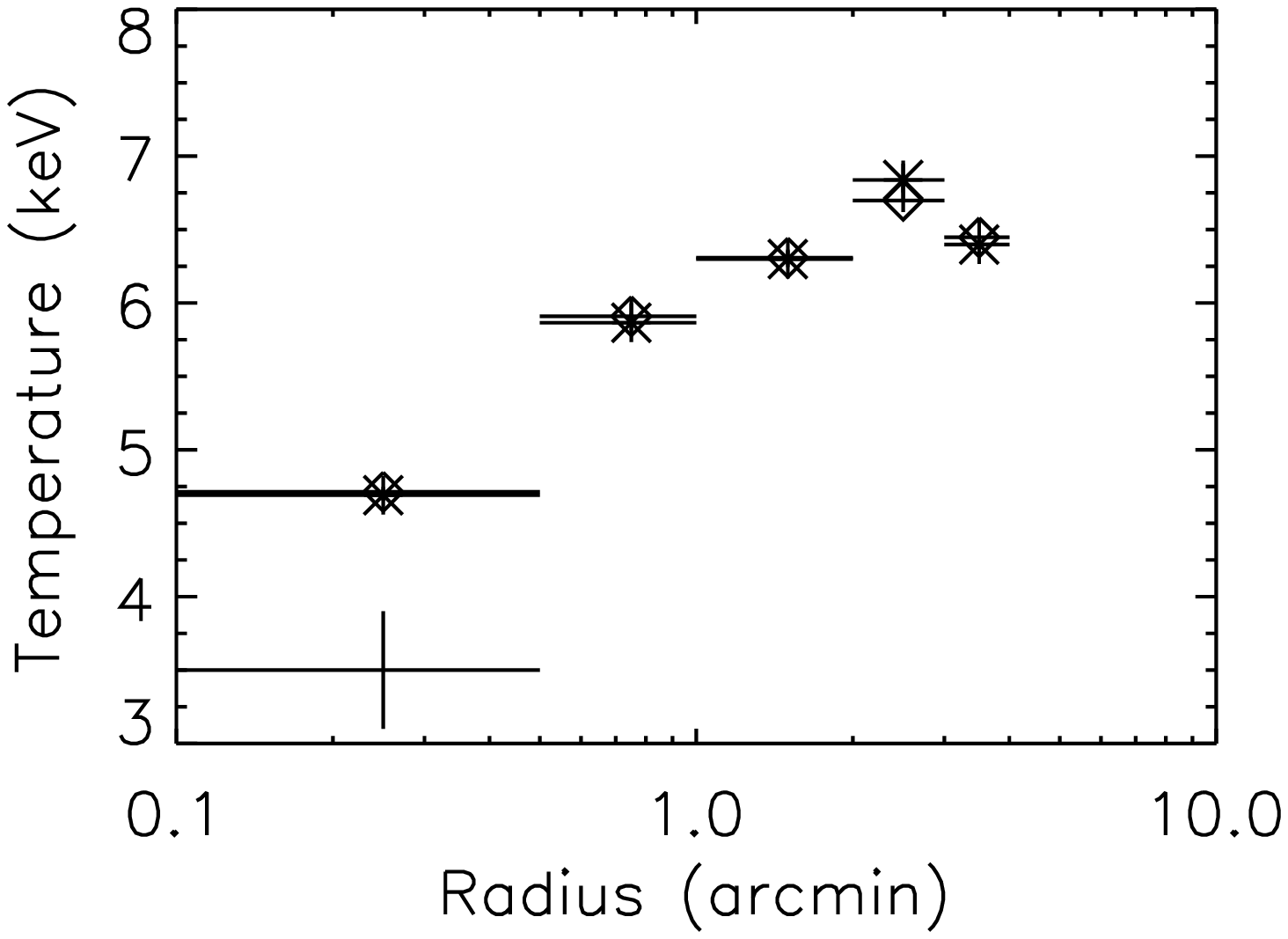}
\end{minipage}
\begin{minipage}{0.33\textwidth}
\includegraphics[width=1.0\textwidth]{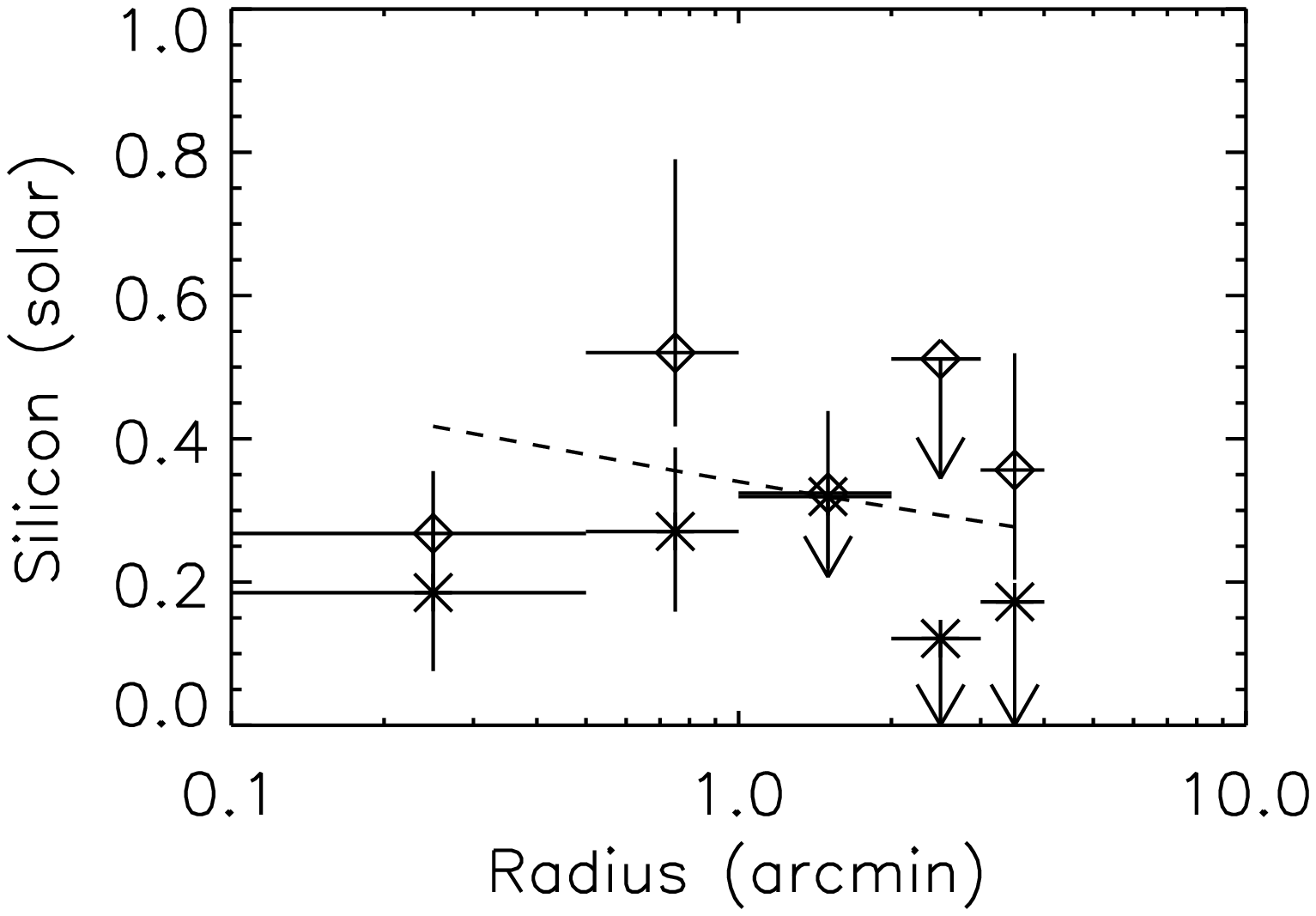}
\end{minipage}
\begin{minipage}{0.33\textwidth}
\includegraphics[width=1.0\textwidth]{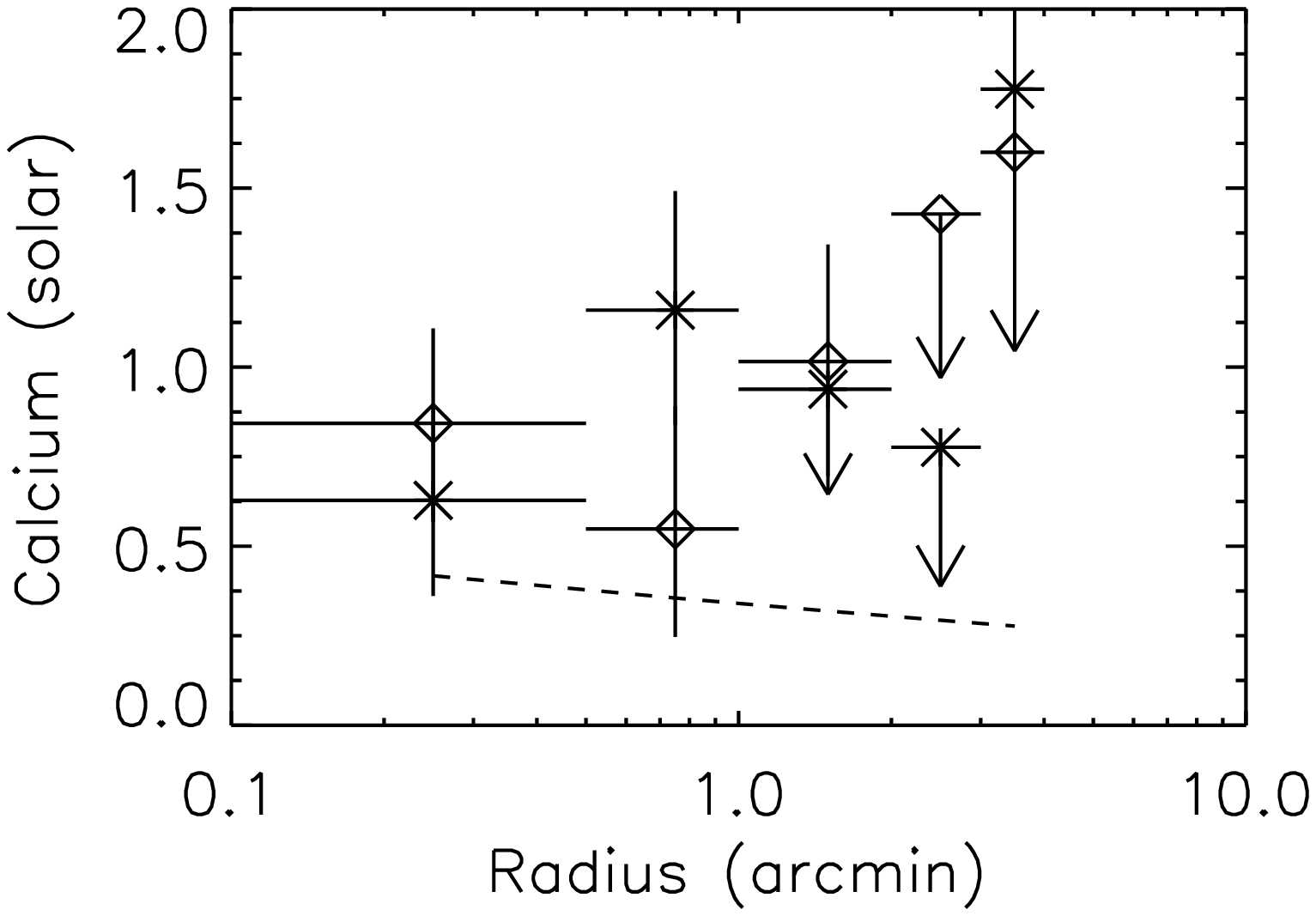}
\end{minipage}\\
\vspace{2mm}

\begin{minipage}{0.33\textwidth}
\includegraphics[width=1.0\textwidth]{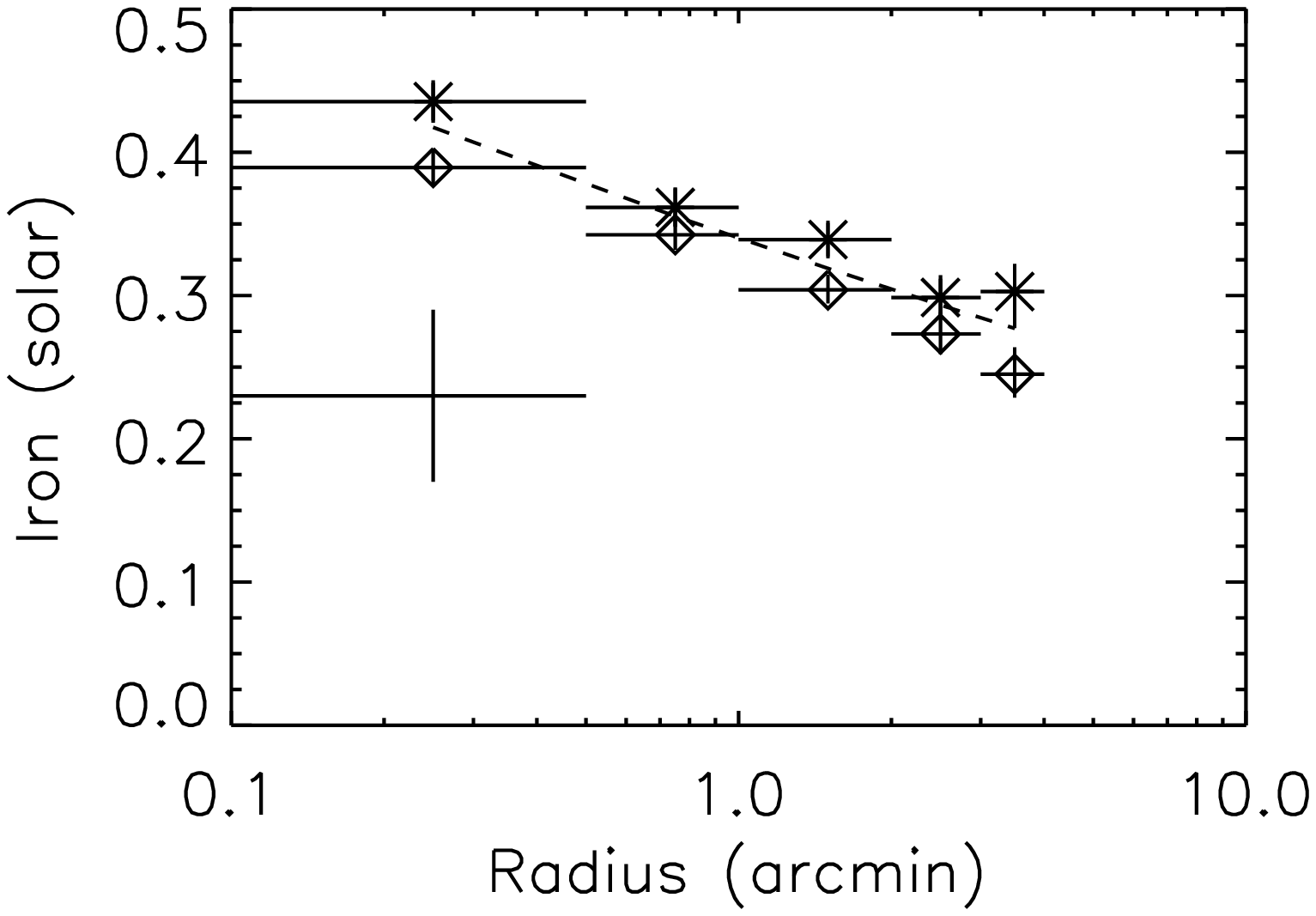}
\end{minipage}
\begin{minipage}{0.33\textwidth}
\includegraphics[width=1.0\textwidth]{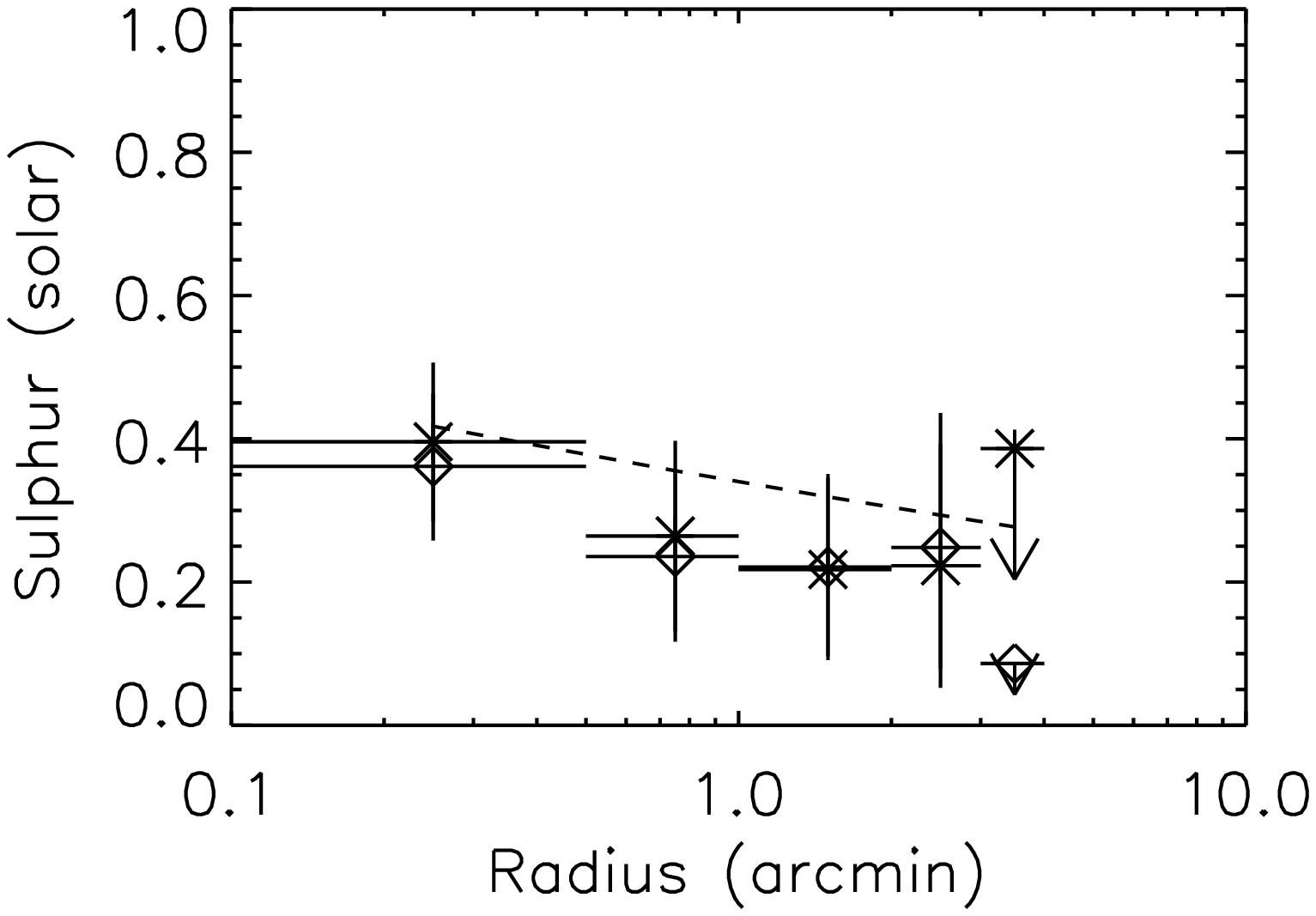}
\end{minipage}
\begin{minipage}{0.33\textwidth}
\includegraphics[width=1.0\textwidth]{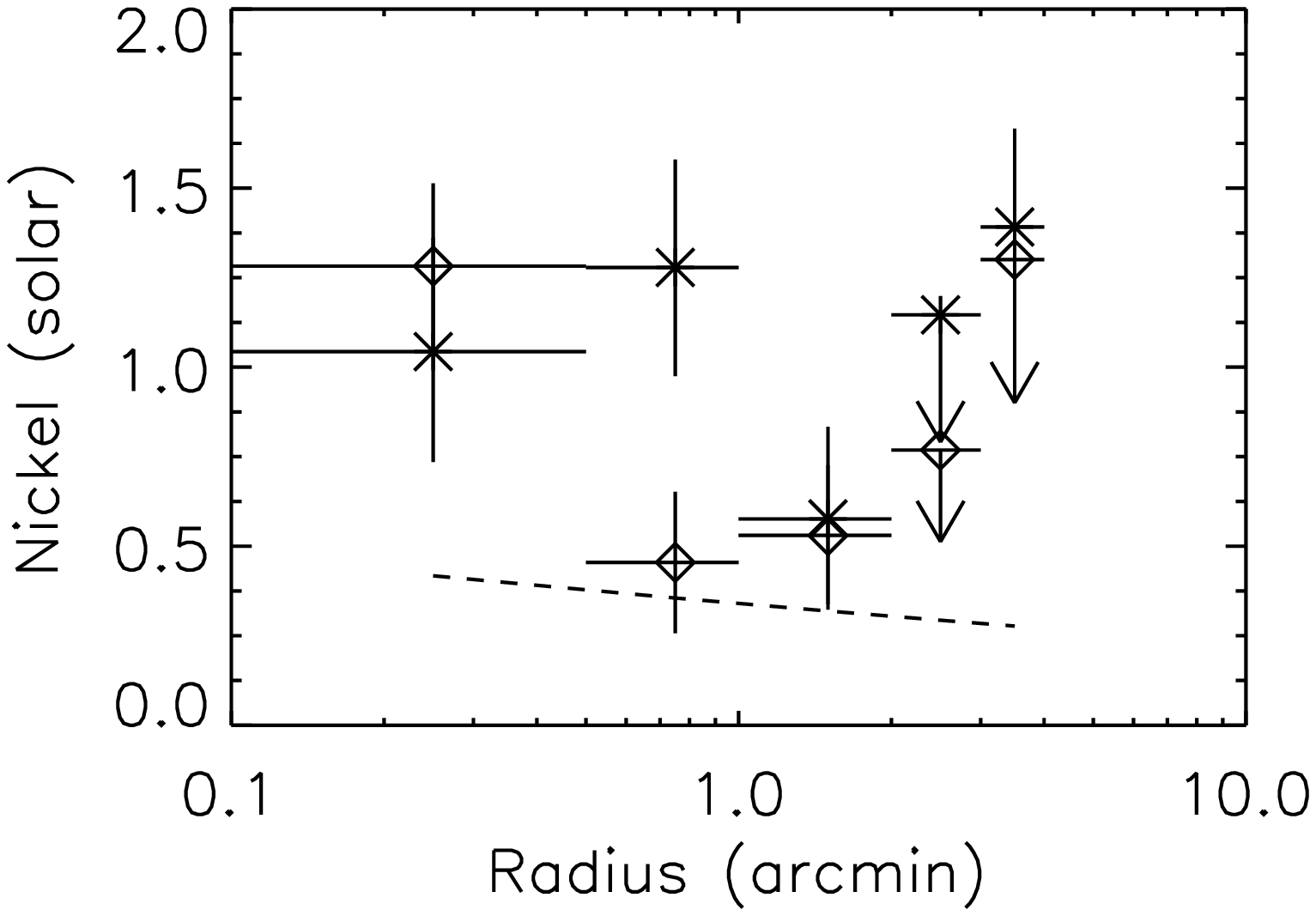}
\end{minipage}
\caption{Single-temperature fit results. EPIC MOS (stars), pn (diamonds) and RGS (plus-sign) are shown. The 
dashed line shows an empirical powerlaw fit to the Fe abundance distribution for comparison.}
\label{fig:singletemp}
\end{figure*}

\subsection{RGS}

We extract the RGS spectra with SAS version 5.4.1 following the same method as described in \citet{tamura2001b}. 
We select the events from a rectangular area on the CCD strip with a full width of 1$\arcmin$ in the cross-dispersion 
direction and centered on the core of the source. We extract the background spectra from several blank field observations \citep{tamura2003a}. 
Because the observation was performed before the cooling of the RGS instruments, the spectra contain a relatively
high number of bad columns due to warm pixels. These bad columns appear as absorption-like features in the spectra. 
Unfortunately, also CCD gaps are present in the wavelength ranges where line 
emission is expected. Together with the relatively high thermal Bremsstrahlung component due to the high temperature 
of the cluster, the lines are difficult to resolve. Therefore, we also include the second order spectrum in the fit to 
increase the statistics. All RGS spectra are fitted over the 8--35 \AA~range.

Because the RGS gratings operate without a slit, the resulting spectrum of an extended source is the sum of
all spectra in the (in our case) $1\arcmin \times \sim12\arcmin$ field of view, convolved with the PSF 
\citep[see for a complete discussion about grating responses][]{davis2001}. 
Extended line-emission appears to be broadened depending on the spatial extent of the source along the dispersion direction. 
In order to describe the data properly, the spectral fits need to account for this effect. In practice, this is accomplished by
convolving the spectral models with the surface brightness profile of the source along the dispersion direction \citep{tamura2004}.
For that purpose we extract the cluster intensity profile from MOS1 along the dispersion direction of RGS, 
which we convolve with the RGS response during spectral fitting. 
Because the radial profile of an ion can be different from the mean profile, this method is not ideal. We 
let the scale of the width and the position of the profile free in the fit to match the profiles of the main 
emission lines.  

\begin{table*}[t]
\caption{
Averaged EPIC results for Abell 478 for the single temperature (1) and {\it wdem} (2) fits. 
Errors are given at the 1$\sigma$ confidence level and the $<$-sign denotes a 2$\sigma$ upper limit.
We leave the abundance of oxygen in the absorption component free to fit the oxygen edge near 0.5 keV. 
This parameter is called $O_{\mathrm{abs}}$. The abundances are given relative to solar abundances measured by \citet{anders1989}. 
}
\begin{center}
\begin{tabular}{l|c|ccccc}
\hline\hline
			& Model	& \multicolumn{5}{c}{EPIC}	 							 		 	 \\
			& 	& 0--0.5$\arcmin$	& 0.5--1.0$\arcmin$	& 1.0--2.0$\arcmin$	& 2.0--3.0$\arcmin$	 & 3.0--4.0$\arcmin$	\\
\hline
$\chi^2$/d.o.f. (MOS)	& 1	& 318 / 327		& 286 / 327		& 290 / 327		& 288 / 327		 & 281 / 303		\\
			& 2	& 247 / 326		& 285 / 326		& 292 / 326		& 278 / 326		 & 281 / 302		\\
\hline
$\chi^2$/d.o.f. (pn)	& 1	& 216 / 179 		& 204 / 179		& 228 / 179		& 211 / 179		 & 166 / 158    	\\
			& 2 	& 177 / 178 		& 193 / 178		& 228 / 178		& 209 / 178		 & 161 / 157    	\\
\hline
$N_{\mathrm{H}}$ 
(10$^{21}$ cm$^{-2}$)	& 1	& 3.54 $\pm$ 0.06	& 3.40 $\pm$ 0.05	& 3.32 $\pm$ 0.03	& 3.34 $\pm$ 0.06	 & 3.49 $\pm$ 0.11	\\
			& 2	& 3.60 $\pm$ 0.04	& 3.44 $\pm$ 0.04	& 3.32 $\pm$ 0.04	& 3.34 $\pm$ 0.05	 & 3.52 $\pm$ 0.05	\\
\hline
O$_{\mathrm{abs}}$	& 1	& 0.40 $\pm$ 0.02	& 0.44 $\pm$ 0.03	& 0.41 $\pm$ 0.02	& 0.39 $\pm$ 0.04	 & 0.32 $\pm$ 0.05 	\\
			& 2 	& 0.50 $\pm$ 0.02	& 0.45 $\pm$ 0.02	& 0.43 $\pm$ 0.03	& 0.38 $\pm$ 0.03	 & 0.34 $\pm$ 0.06 	\\
\hline
$kT$ (keV)		& 1	& 4.70 $\pm$ 0.03	& 5.88 $\pm$ 0.04	& 6.30 $\pm$ 0.04	& 6.76 $\pm$ 0.07  	 & 6.42 $\pm$ 0.07 	\\
\hline
$kT_{\mathrm{max}}$ (keV)& 2	& 8.31 $\pm$ 0.08	& 7.7 $\pm$ 0.3		& 7.0 $\pm$ 0.4		& 8.8 $\pm$ 0.4		 & 9.5 $\pm$ 0.5			\\
\hline
$1/\alpha$		& 2	& 1.12$^{+0.12}_{-0.3}$	& 0.39 $\pm$ 0.15	& 0.12 $\pm$ 0.03	& 0.40 $\pm$ 0.11	 & 0.8 $\pm$ 0.4	\\
\hline
O			& 1	& $<$ 0.06		& $<$ 0.14		& $<$ 0.09		& $<$ 0.09		 & $<$ 0.11  	   	\\
			& 2 	& $<$ 0.07		& $<$ 0.06		& $<$ 0.09		& $<$ 0.06		 & $<$ 0.05  	   	\\
\hline
Ne			& 1	& 1.32 $\pm$ 0.11	& 0.90 $\pm$ 0.12	& 0.63 $\pm$ 0.11	& $<$ 0.5		 & 1.04 $\pm$ 0.17	\\
			& 2 	& $<$ 0.18		& 0.52 $\pm$ 0.11	& 0.60 $\pm$ 0.11	& 0.4 $\pm$ 0.2		 & 0.43 $\pm$ 0.16	\\
\hline
Mg			& 1 	& 0.49 $\pm$ 0.19	& 0.63 $\pm$ 0.15	& 0.51 $\pm$ 0.15	& $<$ 0.5	         & 0.53 $\pm$ 0.19	\\
			& 2	& $<$ 0.08		& 0.46 $\pm$ 0.14	& 0.49 $\pm$ 0.15	& $<$ 0.4      		 & 0.31 $\pm$ 0.17	\\
\hline
Si			& 1	& 0.23 $\pm$ 0.07	& 0.40 $\pm$ 0.08	& 0.20 $\pm$ 0.07	& $<$ 0.5	         & 0.36 $\pm$ 0.16		\\
			& 2 	& 0.22 $\pm$ 0.06	& 0.35 $\pm$ 0.07	& 0.20 $\pm$ 0.07	& $<$ 0.4	         & 0.32 $\pm$ 0.13	\\
\hline
S			& 1	& 0.38 $\pm$ 0.08	& 0.25 $\pm$ 0.09	& 0.22 $\pm$ 0.09	& 0.24 $\pm$ 0.12        & $<$ 0.08		\\
			& 2	& 0.41 $\pm$ 0.06	& 0.25 $\pm$ 0.09	& 0.22 $\pm$ 0.09	& 0.23 $\pm$ 0.11        & $<$ 0.09		\\
\hline
Ar			& 1	& $<$ 0.4		& $<$ 0.2		& $<$ 0.2		& $<$ 0.15		 & $<$ 0.6		\\
			& 2 	& 0.35 $\pm$ 0.15	& $<$ 0.2		& $<$ 0.2		& $<$ 0.16		 & $<$ 0.6		\\
\hline
Ca			& 1	& 0.73 $\pm$ 0.19	& 0.8 $\pm$ 0.2		& 0.7 $\pm$ 0.2		& $<$ 0.8		 & 1.1 $\pm$ 0.4	\\
			& 2 	& 1.1 $\pm$ 0.2		& 1.0 $\pm$ 0.2		& 0.7 $\pm$ 0.2		& $<$ 0.9		 & 1.2 $\pm$ 0.4	\\
\hline
Fe 			& 1	& 0.410 $\pm$ 0.010     & 0.351 $\pm$ 0.009	& 0.320 $\pm$ 0.008	& 0.285 $\pm$ 0.011	 & 0.268 $\pm$ 0.014	\\
 			& 2 	& 0.425 $\pm$ 0.010     & 0.367 $\pm$ 0.009	& 0.322 $\pm$ 0.008	& 0.294 $\pm$ 0.011	 & 0.292 $\pm$ 0.014	\\
\hline
Ni			& 1	& 1.2 $\pm$ 0.2		& 0.78 $\pm$ 0.19	& 0.55 $\pm$ 0.17	& $<$ 0.8	         & 0.9 $\pm$ 0.3 	\\
			& 2 	& 0.79 $\pm$ 0.19	& 0.7  $\pm$ 0.2	& 0.54 $\pm$ 0.17	& $<$ 0.8	         & 0.7 $\pm$ 0.3 	\\
\hline
\end{tabular}
\end{center}
\label{tab:epic_results}
\end{table*}

\subsection{Spectral models}
\label{sec:wdem}

We fit the spectra both with a single temperature collisionally ionized plasma model (MEKAL) and a differential emission measure (DEM) model 
called {\it wdem}. In this particular model the emission measure ($Y$) is distributed as a function of temperature ($T$) 
as shown in Eq.~(\ref{eq:dy_dt}) adapted from \citet{kaastra2004}:
\begin{equation}
\frac{dY}{dT} = \left\{ \begin{array}{ll}
cT^{\alpha} & T < T_{\mathrm{max}} \\
0 & T > T_{\mathrm{max}} \\
\end{array} \right.
\label{eq:dy_dt}
\end{equation}
This model is an empirical parametrization of the DEM distribution found in the core of many clusters. 
In this form the limit $\alpha \to \infty$ yields the isothermal model. For convenience we will use $1/\alpha$ in this paper, 
because then the isothermal model is obtained when $1/\alpha = 0$.
The classical cooling-flow model corresponds to a $dY/dT$ of 1/$\Lambda(T)$, where $\Lambda(T)$ is the cooling function.

To investigate the role of projection effects, we fit the spectra of the core also with an extra temperature component of 6.5 keV.
This way we emulate the spectral contribution of the outer parts of the cluster. 

We notice from a preliminary analysis that the oxygen edge near 0.5 keV is not well fitted. Therefore, we use 
an absorption model component with variable element abundances and leave the oxygen abundance free. We call
this particular oxygen abundance O$_{\mathrm{abs}}$ further on in the paper. 

\section{Results}

\subsection{EPIC}

The results of the EPIC single-temperature fits are presented in Fig.~\ref{fig:singletemp} and Table~\ref{tab:epic_results}. 
The upper left plot in Fig.~\ref{fig:singletemp} shows that the absorption is not a constant across the cluster. 
The temperature profile shows a steep decrease toward the core with a hint for a bend outside the 2.0$\arcmin$ radius. The abundances 
of neon, sulfur and iron are consistent with an increase toward the center. However, neon and sulfur are also consistent with a flat 
distribution, like the other abundances. Because of low statistics we had to discard the 
8--10 keV band in the 3-4$\arcmin$ annulus.

\begin{table}[t]
\caption{Averaged results from the single-temperature (3) and {\it wdem} (4) fits with an extra 6.5 keV temperature component.}
\begin{center}
\begin{tabular}{l|c|cc}
\hline\hline
			& Model	&	\multicolumn{2}{c}{EPIC} \\
			&	& 0--0.5$\arcmin$ & 0.5--1.0$\arcmin$ \\
\hline
$\chi^2$/d.o.f. (MOS)	& 3	& 244 / 326	& 284 / 326	\\
			& 4 	& 247 / 325	& 282 / 325	\\
\hline	
$\chi^2$/d.o.f. (pn)    & 3	& 178 /178	& 192 / 178	\\
			& 4	& 176 / 177	& 200 / 177	\\
\hline
$N_{\mathrm{H}}$ 
(10$^{21}$ cm$^{-2}$) 	& 3	& 3.60 $\pm$ 0.03	& 3.45 $\pm$ 0.06 \\
			& 4	& 3.58 $\pm$ 0.03	& 3.43 $\pm$ 0.06 \\
\hline
O$_{\mathrm{abs}}$	& 3	& 0.453 $\pm$ 0.016	& 0.41 $\pm$ 0.04 \\
			& 4	& 0.488 $\pm$ 0.017	& 0.48 $\pm$ 0.03 \\
\hline
$kT$ (keV)		& 3	& 2.46 $\pm$ 0.04	& 4.14 $\pm$ 0.08 \\
\hline
$kT_{\mathrm{max}}$ (keV)& 4	& 3.04 $\pm$ 0.05	& 4.95 $\pm$ 0.17 \\
\hline
1/$\alpha$		& 4	& 0.38 $\pm$ 0.07	& 1.0 $\pm$ 0.3 \\
\hline			
O			& 3	& $<$ 0.08		& $<$ 0.12 \\
			& 4	& $<$ 0.08		& $<$ 0.05 \\
\hline			
Ne			& 3	& 0.83 $\pm$ 0.16 	& 0.65 $\pm$ 0.14 \\
			& 4	& 0.26 $\pm$ 0.12	& 0.20 $\pm$ 0.10 \\
\hline			
Mg			& 3	& 0.27 $\pm$ 0.07	& 0.50 $\pm$ 0.15 \\
			& 4	& $<$ 0.09		& 0.43 $\pm$ 0.13 \\
\hline			
Si			& 3	& 0.23 $\pm$ 0.05	& 0.36 $\pm$ 0.08 \\
			& 4	& 0.22 $\pm$ 0.05	& 0.33 $\pm$ 0.06 \\
\hline			
S			& 3	& 0.38 $\pm$ 0.06	& 0.27 $\pm$ 0.09 \\
			& 4	& 0.39 $\pm$ 0.06	& 0.26 $\pm$ 0.08 \\
\hline			
Ar			& 3	& 0.31 $\pm$ 0.14	& $<$ 0.18	\\
			& 4	& 0.34 $\pm$ 0.15	& $<$ 0.2	\\
\hline			
Ca			& 3	& 1.1 $\pm$ 0.2		& 1.0 $\pm$ 0.2 \\ 
			& 4	& 1.09 $\pm$ 0.19	& 1.0 $\pm$ 0.2	\\
\hline			
Fe			& 3	& 0.436 $\pm$ 0.010	& 0.362 $\pm$ 0.009 \\
			& 4	& 0.429 $\pm$ 0.010 	& 0.363 $\pm$ 0.009 \\
\hline			
Ni			& 3	& 0.83 $\pm$ 0.18	& 0.7 $\pm$ 0.2		\\
			& 4	& 0.77 $\pm$ 0.19	& 0.69 $\pm$ 0.19	\\
\hline			
\end{tabular}
\end{center}
\label{tab:model-1t}
\end{table}

\begin{figure*}[t]
\centering
\includegraphics[angle=-90,width=17cm]{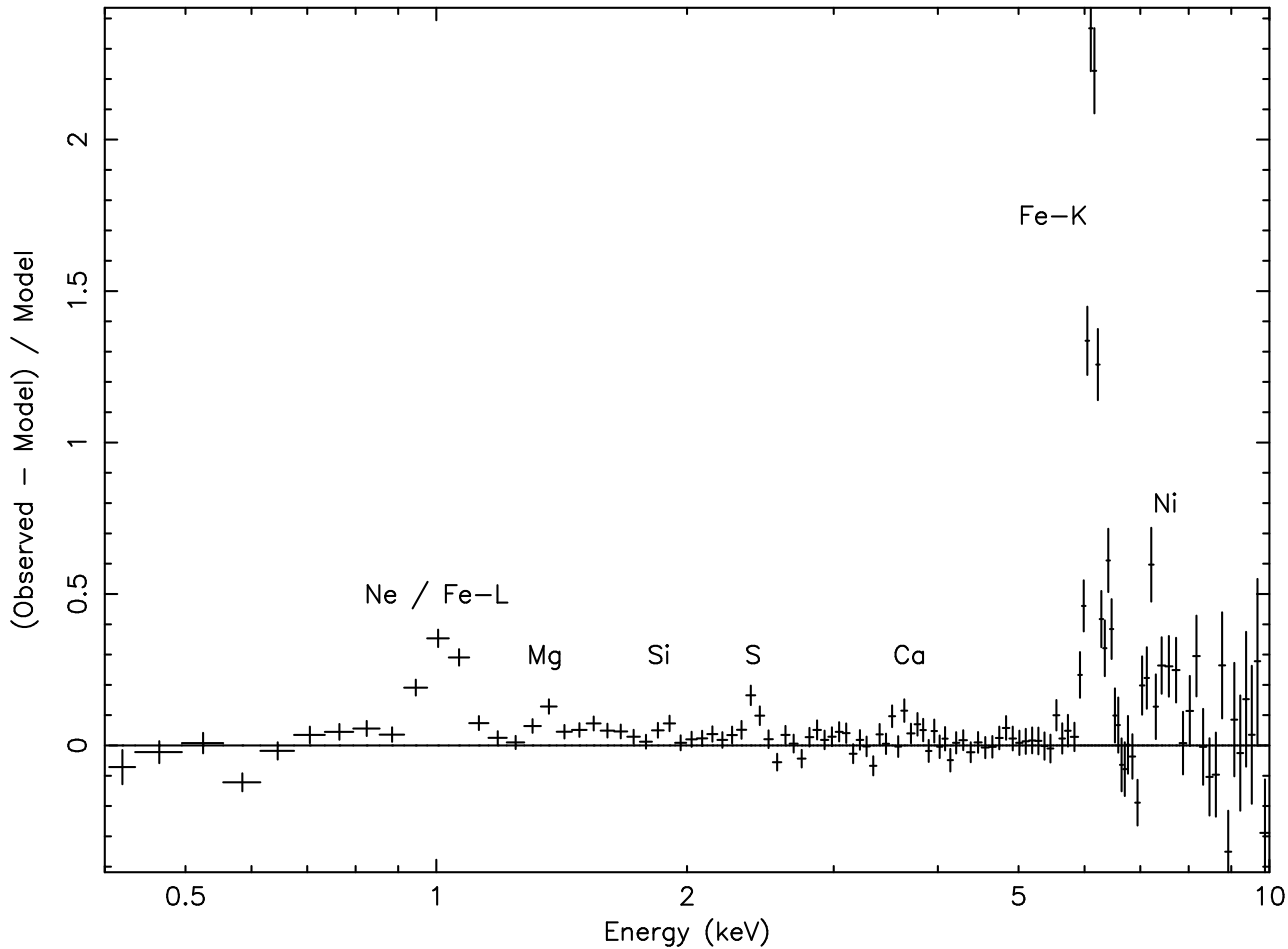}
\caption{Residuals of the averaged MOS spectrum of the core of Abell 478 with the abundances put to zero. }
\label{fig:epic}
\end{figure*}

Fig.~\ref{fig:epic} shows the residuals of the combined MOS spectra of the 0--0.5$\arcmin$ region, fitted with a single 
temperature component for which the abundances have been set to zero afterward. In this way we show the lines that are 
detected in the EPIC spectrum. Except for neon, the abundance of all elements can be fitted
independently. Since the main neon line is close to the Fe-L complex blend at $\sim$1 keV, it is difficult to constrain 
the neon abundance with EPIC. Because the Fe-K line is the strongest line, this line mainly determines the iron abundance.     
The strength of the Fe-L complex is mostly dependent on the temperature structure.

In the core of the cluster the $\chi^2$ value from the single-temperature fit is higher than can be expected from
the high signal-to-noise in that area. We therefore also fit the core with the multi-temperature {\it wdem}
component defined in Sect.~\ref{sec:wdem}. These results are also listed in Table~\ref{tab:epic_results}. The {\it wdem}  abundances
are all consistent with the single temperature abundances, apart from neon, which is blended by Fe-L. From the $\chi^2$ values
the multi-temperature model appears to be preferable over the single-temperature model in the 0--0.5$\arcmin$ region.

Then, we add a single temperature component fixed at 6.5 keV to emulate the 
projection effect from the hot gas in front of the core. The value of 6.5 keV is roughly the mean temperature outside 
the 1$\arcmin$ radius. In Table~\ref{tab:model-1t} we show the results of this fit. The abundances are again consistent
apart from neon. We notice that the fitted temperatures for the core are now significantly cooler than without the hot component.
The value for 1/$\alpha$ in the {\it wdem} component is 0.38 $\pm$ 0.07, lower than the value we previously found, 
1.12$^{+0.12}_{-0.3}$.

\subsection{RGS}

\begin{figure*}[t]
\centering
\includegraphics[angle=-90,width=17cm]{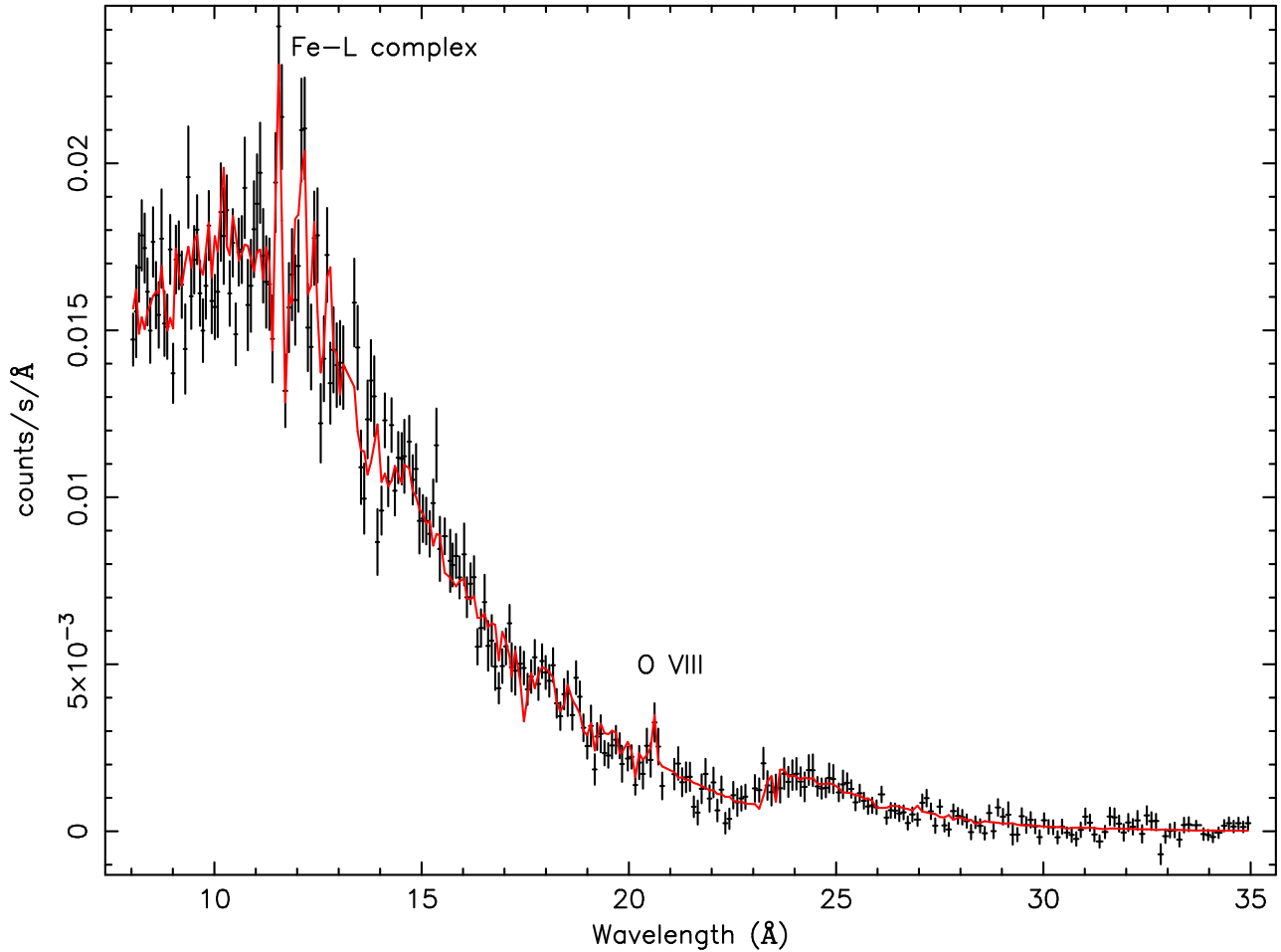}
\caption{RGS averaged spectrum (0--0.5$\arcmin$) of Abell 478. Only the first order spectrum is shown for clarity.
Data points from bad columns were removed from the RGS1 and RGS2 spectra before they were averaged.}
\label{fig:rgs12}
\end{figure*}

\begin{table}[t]
\caption{Results from the RGS spectral fits of the core region (0--0.5$\arcmin$) of Abell 478.
The RGS spectra were fitted using the {\it wdem} model component and a single-temperature model.
Errors are given at the 1$\sigma$ confidence level and the $<$-sign denotes a 2$\sigma$ upper limit.
We left the abundance of oxygen in the absorption component free to fit the oxygen edge near 0.5 keV. 
This parameter is called O$_{\mathrm{abs}}$. 
}
\begin{center}
\begin{tabular}{lccl}
\hline\hline
			& RGS			& RGS 			 & Unit \\
			& wdem			& single-temperature	 & \\
\hline
$\chi^2$/d.o.f		& 867 / 753		& 887 / 754		 & 	\\
$N_{\mathrm{H}}$	& 3.15 $\pm$ 0.08	& 3.30 $\pm$ 0.07	 & 10$^{21}$ cm$^{-2}$ \\
O$_{\mathrm{abs}}$	& 0.67 $\pm$ 0.04	& 0.63 $\pm$ 0.04	 & Z$_{\sun}$ \\
$kT$			&			& 3.5 $\pm$ 0.4		 & keV \\
$kT_{\mathrm{max}}$	& 11.1$^{+1.2}_{-2.3}$	& 			 & keV \\
1/$\alpha$		& 1.3 $\pm$ 0.3		& 			 & \\
O			& 0.15 $\pm$ 0.07	& 0.09 $\pm$ 0.05	 & Z$_{\sun}$ \\
Ne			& 0.46 $\pm$ 0.13	& 0.39 $\pm$ 0.10	 & Z$_{\sun}$ \\
Mg			& 0.17 $\pm$ 0.12	& $<$ 0.2		 & Z$_{\sun}$ \\
Fe 			& 0.61$^{+0.10}_{-0.17}$& 0.23 $\pm$ 0.06	 & Z$_{\sun}$ \\
\hline
\end{tabular}
\end{center}
\label{tab:rgs_results}
\end{table}

We obtain a reasonable fit to the RGS spectra of the core (0--0.5$\arcmin$) with the single-temperature 
model. The $\chi^2$ / d.o.f. is 887 / 754. Unfortunately, the \ion{O}{viii} line complex is partly within 
a CCD bad-column on RGS1 and falls in the dead area of RGS2, which makes the derived oxygen abundance more uncertain.
The Fe-L complex and Ne line near 12 \AA~are detected and resolved. 

Because the EPIC spectra from the core region are best fitted with a multi-temperature model, we also fit the
RGS spectra with the {\it wdem} component. This results in a slightly better $\chi^2$ / d.o.f. of 867 / 753. The results
of this fit and the single-temperature model are shown in Table~\ref{tab:rgs_results} and Fig.~\ref{fig:rgs12}. 
Because the temperature determination with RGS depends mostly on the line emission, the weakness of the lines in 
this spectrum causes the temperature to be less well constrained. Furthermore, the energy range of the RGS is 
much smaller than EPIC and concentrated on the soft energy part of the spectrum, making it more sensitive 
to the cool component of the spectrum. Therefore, the fit is less sensitive to the multi-temperature distribution. Because  
of the increased freedom of the fit due to the use of multi-temperature components, the {\it wdem} fit is marginally 
better than the single-temperature fit. We have also attempted to fit
the RGS spectrum with an extra hot component of 6.5 keV which we added to both a single-temperature and {\it wdem} model. 
Because the temperatures in this model could not be constrained, we have not pursued this model any further.

\section{Discussion}

We analyze the high-resolution spectra of the \object{Abell 478} cluster of galaxies and we derive, for the
first time, radial abundance profiles for several elements in this hot cluster. The exposure time 
of 126 ks allows us to resolve the main emission lines. By fitting these lines and the continuum emission
we are able to put constraints on the temperature structure in the core and the abundance distribution of 
Ne, Mg, Si, S, Ca, Fe and Ni. Furthermore, by fitting the oxygen edge near 0.5 keV we measure an underabundance 
of oxygen in the absorption component.     

The temperature profile ob\-tain\-ed from the single-temperature fits (see Fig.~\ref{fig:singletemp} and 
Table~\ref{tab:epic_results}) of the core ($<$ 2$\arcmin$) are 
consistent with the temperature profile measured with Chandra \citep{sun2003} and ROSAT \citep{white1994,allen1993}. 
It confirms that \object{Abell 478} has one of the steepest temperature gradients observed so far. 

In the outer regions, outside 2$\arcmin$ from the core, our pn temperature profile shows a hint of a bend which is consistent with 
the more extended temperature determination by \citet{pointecouteau2004}, but is inconsistent with \citet{sun2003}. 
The ACIS aboard Chandra, which has a lower sensitivity at high energies and a smaller field of view than pn, 
is therefore more vulnerable to systematic background effects in the temperature. The bend in the pn temperature
profile is comparable with profiles observed in other clusters, for example the core of S\'ersic 159-03 
\citep{kaastra2001}, and has been interpreted as the transition from the cooling core to the inter-cluster medium.    

In the core region, within 0.5$\arcmin$, a multi-temperature ({\it wdem}) model fits better than the 
single-temperature model. The value of 1/$\alpha$ = 1.3 $\pm$ 0.3 from RGS is high compared to the cluster samples studied 
by e.g. \citet{peterson2003} and \citet{kaastra2004}. These studies find values for 1/$\alpha$ which are 
roughly of the order of 0.5. If we add an extra temperature component of 6.5 keV to account for the emission from the outer 
regions in front of the core, then the difference between the single-temperature and {\it wdem} component vanishes (see Table~\ref{tab:model-1t}). 
This suggests that the projection effect of the hot cluster material along the line of sight strongly affects the 1/$\alpha$ value
in the fits. 

The multi-temperature behavior is not necessarily explained by contamination from the outer parts of the cluster. 
Although the X-ray cavities reported by \citet{sun2003} are not spatially resolved with 
XMM-Newton, the value of 1/$\alpha$ and the cool area associated with the X-ray cavity may be related.
Together with the high resolution study of the temperature profile \citep{pointecouteau2004}, which
reveals a steep temperature gradient in the core region, these effects could also partially account for the multi-temperature 
behavior observed in the central bin. Unfortunately, the spatial resolution of XMM-Newton does not allow
us to draw conclusions from these data on the complex structure in the core.

From the single-temperature fits we measure a central increase of the iron abundance from $\sim$0.3 at about 4$\arcmin$ 
to $\sim$0.4 in the center which is consistent with the mean abundance profile of Abell 478 derived by \citet{white2000} using ASCA. 
The central abundance is also consistent with earlier EXOSAT measurements by \citet{edge1991}. 
The significant central increase of Fe is similar to the profiles of other hot ($>$ 6 keV) clusters \citep{tamura2004}. 
Also neon and sulfur seem to follow the same trend in our results. However, due to the large error bars these profiles could still 
be consistent with a flat distribution. The central increase of sulfur was also found in the cluster sample of \citet{tamura2004}.  
Neon is not resolved in EPIC, because of the lower spectral resolution and blending with lines from the iron L complex.
Therefore, the systematic uncertainty on the abundance of neon is quite large. Moreover, the single-temperature
model tries to fit the Fe-L complex by enhancing the Ne abundance, while the reason for the Fe-L to be enhanced is the 
presence of colder material. Because RGS resolves the neon line, its value for the neon abundance is much more robust. 
Neon and oxygen are thought to have the same origin. Therefore, we expect their values and spatial behavior
to be comparable, which excludes the very high numbers for neon from the single-temperature EPIC fits. 
The nickel abundance we find is likely to be overestimated due to an error in the nickel line 
energies \citep[see also][]{gastaldello2004}. 

The oxygen abundance could not be constrained in our EPIC fits, but despite the occurrence of bad columns in the line 
resolved by the RGS we could get a value for the oxygen abundance consistent with the EPIC upper limits.  
We can compare our result with the predicted O/Fe ratios from theoretical models for supernovae Ia and II. 
The O/Fe ratio obtained from the RGS {\it wdem} fit is 0.25 $\pm$ 0.13, which is lower than the average value
of 1.2 $\pm$ 1.6 for hot clusters reported by \citet{tamura2004}, but not significantly different because of the large 
spread. The theoretical models predict an O/Fe ratio of $<$ 0.05 and 1.5--4 for supernova types Ia and II, respectively.
Our value of 0.25 $\pm$ 0.13 is in between the two predictions, like in the other hot clusters.

\begin{figure}[t]
\resizebox{\hsize}{!}{\includegraphics{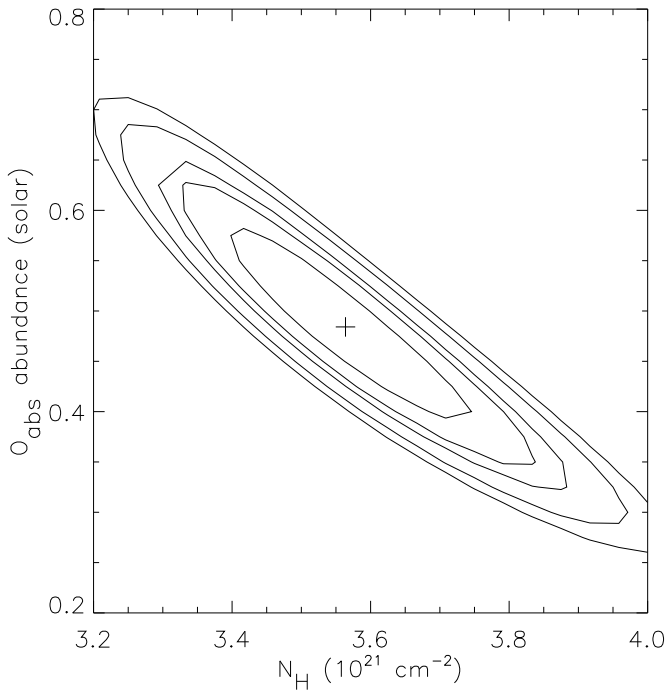}}
\caption{Error ellipses for 
$N_{\mathrm{H}}$ against O$_{\mathrm{abs}}$ calculated from the MOS fit of the core (0.0--0.5$\arcmin$) region. 
Contours are drawn for $\Delta\chi^2 =$ [2.30, 4.61, 6.71, 9.21, 11.8].}
\label{fig:nh_oabs_contour}
\end{figure}

In Table~\ref{tab:epic_results} and Table~\ref{tab:rgs_results} the values for $N_{\mathrm{H}}$ and O$_{\mathrm{abs}}$ appear to be  
different between EPIC and RGS. This is partly due to the fact that the EPIC/RGS cross-calibration near the oxygen edge 
is still not optimal. The EPIC results are also slightly affected by correlations between $N_{\mathrm{H}}$, O$_{\mathrm{abs}}$ and the oxygen abundance,
because of blending due to the lower CCD resolution.
The O$_{\mathrm{abs}}$ anti-correlates with absorption, but because fixing the edge results  
in large residuals at lower energies, we decided to let the oxygen abundance in the absorption component free. 
Fig.~\ref{fig:nh_oabs_contour} shows the error ellipses for the oxygen abundance against N$_{\mathrm{H}}$. 
From the plot it is apparent that N$_{\mathrm{H}}$ and O$_{\mathrm{abs}}$ are anti-correlated.
If we fix the O$_{\mathrm{abs}}$ abundance to 1.0, the N$_{\mathrm{H}}$ values we find are consistent with those from
\citet{pointecouteau2004} and \citet{sun2003}. This plot also shows that
O$_{\mathrm{abs}}$ is significantly different from solar. \citet{weisskopf2004} report from their Chandra observation of the 
\object{Crab} nebula that, using the abundances from \citet{anders1989}, they find an underabundance of oxygen in the Galactic absorption component
of 0.41 $\pm$ 0.07. \citet{takei2003} find values in the range from 0.63 $\pm$ 0.12 to 0.74 $\pm$ 0.14 using a Chandra 
observation of \object{Cyg X-2}. Within the calibration uncertainties our results are consistent with these numbers.
It is likely that the observed underabundance is an artifact of the solar abundances we use. \citet{allende2001}
reported a new solar photospheric abundance of log $\epsilon({\mathrm{O}}) = 8.69 \pm 0.05$ dex. This is 
0.58 $\pm$ 0.08 times smaller than the value of \citet{anders1989} and in between our values for O$_{\mathrm{abs}}$.
\citet{pointecouteau2004} show that the origin of the absorption in the direction of \object{Abell 478} is 
most likely Galactic and correlated with infra-red data.
Although O$_{\mathrm{abs}}$ is consistent with the solar oxygen abundance of \citet{allende2001}
we cannot fully exclude the existence of a real underabundance of oxygen in the absorbing material along the line of sight. 

Finally, we find that the average redshift measured in the pn detector is 0.0775 $\pm$ 0.0002, compared to 0.0881 $\pm$ 0.0009
measured in the optical by \citet{zabludoff1990}. This is probably a gain-related problem which was not
corrected for during the XMM-Newton SOC data processing. Problems with the redshift determinations are not unique.
Discrepancies between X-ray measured redshifts and optical redshifts have been reported earlier
by \citet{zhang2004} and \citet{takahashi2003}. 

\section{Conclusions}

We analyze the high-resolution XMM-Newton spectra of the Abell 478 clusters of galaxies and conclude that:

\begin{itemize}
\item We measure an underabundance of oxygen in the Galactic absorption component. This is measured for the first time using 
an observation of a cluster of galaxies.  
\item We derive radial abundance profiles for Ne, Mg, Si, S, Ca, Fe and Ni, which confirm the trends observed in
other clusters \citep{tamura2004}.
\item The core of the cluster shows multi-temperature behavior, which is mostly explained by projection effects
partly because of the steep temperature gradient.  
\end{itemize}

\begin{acknowledgements}
We would like to thank Monique Arnaud for providing useful comments and for the 
fruitful discussions we had while completing this paper. 
This work is based on observations obtained with XMM-Newton, an ESA science
mission with instruments and contributions directly funded by ESA member states
and the USA (NASA). The Space Research Organization of the Netherlands (SRON)
is supported financially by NWO, the Netherlands Organization for Scientific
Research.
\end{acknowledgements}

\bibliographystyle{aa}
\bibliography{clusters}

\end{document}